\documentclass[twocolumn,english,aps,prd,reprint,floatfix,notitlepage,nofootinbib,preprintnumbers,superscriptaddress,longbibliography]{revtex4-1}
\pdfoutput=1
\usepackage{lmodern}

\usepackage[T1]{fontenc}
\usepackage[latin9]{inputenc}
\usepackage{geometry}
\usepackage{subfigure,lmodern, amsmath,amssymb, graphicx, pifont, adjustbox, bm, xcolor}
\usepackage{amsfonts}
\usepackage{enumitem}
\usepackage{comment}
\usepackage{mathtools}
\usepackage{float}
\usepackage{slashed}
\usepackage{ragged2e}
\usepackage{array}
\usepackage{bbm}
\usepackage{balance}
 \usepackage{booktabs}
\usepackage{multirow}
\usepackage{tikz}

\usepackage{nameref}

\usepackage{hhline}

\makeatletter\g@addto@macro\bfseries{\boldmath}\makeatother

\makeatletter\newcommand{\labeltext}[2]{%
  \def\@currentlabel{#1}%
  \label{#2}%
}
\makeatother

\usepackage{stackengine}
\usepackage{esint}
\usepackage[unicode=true,pdfusetitle,
 bookmarks=true,bookmarksnumbered=false,bookmarksopen=false,
 breaklinks=false,pdfborder={0 0 1},backref=false,colorlinks=true]
 {hyperref}
\hypersetup{
 pdfauthor={Clifford Cheung, Rachel Rosen},
 citecolor=blue,linkcolor=black,urlcolor=blue}

\newcommand{\appendixref}[1]{\hyperref[#1]{appendix~\ref{#1}}}
\def\equationautorefname~#1\null{eq.\,(#1)\null}
\usepackage{breakurl}
\usepackage[hang,flushmargin]{footmisc} 
\allowdisplaybreaks
\makeatletter

\usepackage{etoolbox}
\apptocmd{\thebibliography}{\justifying\setlength{\leftskip}{7.4mm}}{}{} 
 
 \usepackage{relsize}
\usepackage{babel}

\makeatletter
\def\simgt{\mathrel{\lower2.5pt\vbox{\lineskip=0pt\baselineskip=0pt
           \hbox{$>$}\hbox{$\sim$}}}}
\def\simlt{\mathrel{\lower2.5pt\vbox{\lineskip=0pt\baselineskip=0pt
           \hbox{$<$}\hbox{$\sim$}}}}
\makeatother

\usepackage{changepage}

\newcommand{\be}{\begin{equation}}
\newcommand{\ee}{\end{equation}}
\newcommand{\bea}{\begin{eqnarray}}
\newcommand{\eea}{\end{eqnarray}}

\newcommand{\Eq}[1]{Eq.~(\ref{#1})}

\newcommand{\eq}[2]{\be\begin{aligned}#1 \label{#2}\end{aligned}\ee}

\newcommand{\mysec}[1]{\noindent {\bf #1.}---}

\newcolumntype{P}[1]{>{\centering\arraybackslash}p{#1}}

\usepackage{fix-cm}

\definecolor{dartmouthgreen}{rgb}{0.05, 0.5, 0.06}

\begin{document}

\preprint{CALT-TH 2026-022}

\title{Thermal Positivity}

\author{Clifford Cheung}
\affiliation{Walter Burke Institute for Theoretical Physics and
Leinweber Forum for Theoretical Physics, California Institute of Technology, Pasadena, CA 91125, USA}
\author{Rachel A. Rosen}
\affiliation{Department of Physics, Carnegie Mellon University, Pittsburgh, PA 15213}

\begin{abstract} 

\noindent We argue that Lorentz invariance and unitarity impose sharp constraints on thermodynamic quantities. By relating thermal vacuum diagrams to  forward scattering amplitudes, we derive an infinite family of sign conditions on finite-temperature observables in perturbative theories of relativistic massless bosons. In particular, we prove that all low-temperature corrections from interactions to the pressure, or equivalently the negative free energy density,  of the form $T^{2D-4+4k}$ with $k>0$ in $D$ spacetime dimensions, are strictly positive.  These positivity conditions are inherited by analogous terms in the entropy density and specific heat. Our results apply to any effective field theory that is free of long-range forces and descends from a weakly coupled ultraviolet completion, in which case higher-loop and higher-multiplicity thermal diagrams are parametrically subleading.

\noindent 
\end{abstract}

\maketitle

\mysec{Introduction}A central tenet of effective field theory is that all that is not forbidden is compulsory.  By this logic, all low-energy interactions permitted by the underlying symmetries should appear with essentially arbitrary couplings. On the other hand, recent developments have articulated an important counterpoint to this lore: much, in fact, is actually forbidden. Tools from the modern scattering amplitudes program have uncovered a treasure trove of so-called positivity bounds that rigorously constrain the signs of the Wilson coefficients parameterizing the strengths of interactions \cite{Pham:1985cr,Adams:2006sv,Nicolis:2009qm,Arkani-Hamed:2020blm,Caron-Huot:2020cmc,Bellazzini:2020cot,Caron-Huot:2021rmr}. A parallel program has also emerged leveraging causality violation to the same aim \cite{Camanho:2014apa,Papallo:2015rna,Hinterbichler:2017qyt,deRham:2021bll}.

The bulk of work on this subject has centered on relativistic quantum field theories at zero temperature. It is then natural to ask whether or how positivity might persist away from this limit. Indeed, finite temperature is a controlled setting in which the microscopic theory is relativistic even though the thermal state is not. 

The divination of signs in thermodynamics has a historical precedent.
It was conjectured long ago that the free energy might be monotonic along any renormalization group flow \cite{Periwal:1994im,Appelquist:1999hr,Appelquist:1999vs}.  If true, such a statement would constitute a generalization of the $a$- and $c$-theorems \cite{Zamolodchikov:1986gt,Cardy:1988cwa,Komargodski:2011vj}, which rigorously quantify the loss of degrees of freedom, to any spacetime dimension.  Unfortunately, known counterexamples to this old proposal exist \cite{Sachdev:1993pr,Klebanov:2011gs}.  Nevertheless, one might naively expect a more modest claim: that, in the vicinity of a free  or infrared conformally invariant theory, higher dimension operators should lower the thermal free energy and thus increase the thermal entropy because microscopic degrees of freedom are integrated out.  For perturbative, tree-level ultraviolet completions this holds for the entropy of black holes \cite{Cheung:2018cwt,Cheung:2019cwi} relevant to the weak gravity conjecture \cite{Arkani-Hamed:2006emk}.  More recently, the broader validity of these signs was  explored systematically across a swath of theories \cite{Fernandez-Sarmiento:2025tpy}.

In this paper, we derive a novel and infinite class of positivity bounds on thermodynamic observables. Our results apply to any theory of relativistic, massless bosons that is perturbative and unitary.  A critical technical assumption is that long-range forces are absent and the ultraviolet completion is weakly coupled, so higher-loop or higher-multiplicity corrections to thermal observables are subleading.

The key insight of this work is a universal relation that is best summarized diagrammatically,
 \vspace{.2cm}
\eq{
\begin{tikzpicture}[thick, scale=0.62]

  \tikzset{
    vertex/.style={
      draw=black,
      fill=gray!20,
      circle,
      minimum size=8pt,
      inner sep=0pt
    }
  }

  \begin{scope}[shift={(0,0)}]
    \draw (0,0.95) circle (0.95);
    \draw (0,-0.95) circle (0.95);
    \node[vertex] at (0,0) {};
    \node[align=center, font=\small] at (0,-3.25) {thermal loop\\diagram};
  \end{scope}

  \node at (2.75,0) {\Large $\simeq$};

  \begin{scope}[shift={(6.4,0)}]
    \draw (-1.65,1.15) -- (0,0);
    \draw (-1.65,-1.15) -- (0,0);
    \draw (0,0) -- (1.65,1.15);
    \draw (0,0) -- (1.65,-1.15);
    \node[vertex] at (0,0) {};
    \node[align=center, font=\small] at (0,-3.25) {forward scattering\\amplitude};
  \end{scope}
\end{tikzpicture}
}{diagram}
This figure depicts the close relationship between a  two-loop vacuum diagram at {\it finite temperature} and a corresponding tree-level scattering amplitude at {\it zero temperature}.    The former dictates the leading interaction-induced correction to the pressure,
\eq{
\Delta P &= 
- \frac{\Delta F }{V}= \frac{1}{2}  \int d\Pi_1  d\Pi_2 \,A((p_1+p_2)^2,0)\, ,
}{DeltaP}
which by extensivity is equal to the negative of the free energy density.  Here we have defined the thermal phase space integral for relativistic massless bosons,
\eq{
d\Pi =  \frac{ d^{D-1} \vec{p}}{(2\pi)^{D-1} 2 |\vec p|}  
\frac{1}{e^{|\vec p|/T}-1} \, ,
}{dPi}
with analogous formulae for fermions and massive particles.
  By construction, each thermal loop in \Eq{diagram} identifies an incoming and outgoing momentum at the vertex.  This projects the vertex onto the forward limit of a zero-temperature on-shell scattering amplitude, $A(s,t)$, evaluated at the squared center-of-mass energy $s=(p_1+p_2)^2$ and vanishing momentum transfer $t=0$.

Conveniently, on-shell scattering amplitudes are bounded by a host of well-known positivity conditions which can be immediately retrofitted onto thermodynamic observables. Consider any $D$-dimensional, unitary theory of relativistic bosons that is free of long-range forces and whose ultraviolet completion is perturbative.  For such theories, the low-temperature expansion of the pressure correction is
\eq{
\Delta P = \sum_k \xi_k T^{2D-4+4k} + \cdots,
}{free_energy}
where the ellipses denote other powers of the temperature that our framework does not constrain.  The primary claim of this work is the thermal positivity bound,
\eq{
\xi_k >0 \qquad \textrm{for all} \qquad k>0 \, .
}{bound}
The entropy density $\Delta s= \partial \Delta P / \partial T$ and specific heat capacity $c_V= T \partial^2 \Delta P/ \partial T^2$ then inherit the very same positive properties in their low-temperature expansions.

A notable feature of these thermal positivity bounds is that they are valid {\it arbitrarily deep} into the low-temperature expansion, which is to say, for $\xi_k$ at arbitrarily large $k$.  In this regard, these constraints precisely mirror those derived in scattering amplitudes, where higher-dimension operators of arbitrarily high degree enjoy robust positivity conditions. An obvious corollary of our results is that the moment bounds on forward scattering amplitudes \cite{Bellazzini:2020cot} translate directly into constraints on the $\xi_k$ across {\it distinct} values of $k$.

The connection between thermodynamics and dynamical scattering has a long history.  Going back nearly a century, it has been known in non-relativistic systems that the second coefficient in the virial expansion of the free energy is related to the two-body scattering phase shift by the Beth-Uhlenbeck formula \cite{Beth:1937zz,Huang:1963},
\be
\Delta F \sim \int_0^\infty dE \, e^{-E/T} \sum_{\ell} (2 \ell +1) \frac{\partial \delta_\ell(E)}{\partial E} \, .
\ee
A small phase shift $\delta_\ell$ can be written, via the partial wave expansion, in terms of the forward scattering amplitude, $
f(k) \simeq  \frac{1}{k} \sum_\ell (2 \ell+1)  \delta_\ell(k)$.  The relativistic generalization of this formula was derived in the remarkable work of Dashen, Ma, and Bernstein \cite{Dashen:1969ep}, where
\eq{
\Delta F=  -\frac{ T}{2\pi i } \int_0^{\infty} dE \, e^{- E/T}\, {\rm Tr}\left(S^\dagger \partial_E S\right) \, .
}{DMB}
While the above formula exhibits many subtleties regarding the forward scattering limit \cite{Schubring:2024yfi,Baratella:2024sax, Baratella:2025fcj,Albert:2026fqj}, the two-to-two scattering contributions are well-defined and equivalent to the standard thermal vacuum diagram in
\Eq{diagram}.

\medskip
\mysec{Free Energy}To begin, let us consider the Euclidean action for a $D$-dimensional quantum field theory,
\eq{
S = S_0 + \Delta S,
}{}
where $S_0$ and  $\Delta S$ encode the free particle propagation and leading perturbative interactions, respectively.   The partition function is then
\eq{
Z = Z_0 \langle e^{-\Delta S} \rangle_0,
}{}
with all expectation values evaluated on the thermal state defined in the free theory.  The free energy $F=-T \log Z$ includes an interaction-induced correction,
\eq{
\Delta F = F - F_0 = T \langle \Delta S \rangle_0 + \cdots,
}{}
which at leading order in perturbation theory is simply the thermal expectation value of the interaction action in the thermal state of the free theory.

To be concrete, consider a theory of real massless scalars interacting via a four-point interaction,
\eq{
\Delta S= \frac{1}{4!} \prod_{i=1}^4 \int_{p_i} \phi(p_i) \delta^D\left(\sum_{i=1}^4 p_i \right) V(p_1,p_2,p_3,p_4) \, ,
}{}
where we have defined the integral
$\int_{p} =  T \sum_{n} \int\frac{ d^{D-1} \vec{p}}{(2\pi)^{D-1}}$
over the Euclidean four-momenta $p = (\omega_n, \vec p)$ for
$\omega_n = 2\pi n T$, where $n$ labels the Matsubara modes.  
The  off-shell Euclidean four-point vertex $V(p_1,p_2,p_3,p_4)$ is related to the scattering amplitude,
\eq{V(p_1,p_2,p_3,p_4) \overset{\substack{\text{Lorentzian}\\[2pt]\text{on-shell}}}{\vphantom{|}\rightarrow}
 -A(p_1,p_2,p_3,p_4) \, ,
}{}
upon Wick rotation to Lorentzian mostly-plus signature and the imposition of on-shell kinematics. 

To compute the thermal expectation values, we Wick contract with the
 thermal two-point function $\langle \phi(p_1)\phi(p_2) \rangle_0 = G(p_1,p_2) = \delta^D(p_1+p_2) \frac{1}{p^2_1}$
where all dot products are in Euclidean.
Contracting these propagators in each channel, 
we obtain the two-loop thermal vacuum diagram,
\eq{
\Delta F = T\langle \Delta S \rangle_0 &=  \frac{1}{8} V\int_{p_1} \int_{p_2} \frac{V(p_1,-p_1,p_2,-p_2)}{p_1^2 {p}_2^2},
}{}
where we have used $\delta^D(0) =  V/T$.  

Importantly, this thermal loop diagram is literally the integral of the on-shell amplitude.  First, any terms in  $V(p_1,-p_1,p_2,-p_2)$  proportional to $p_1^2$ or $p_2^2$ will cancel the corresponding propagator denominator.  These contributions are analytic and integrate to zero.   Physically, this happens because the thermal phase space integral sums over on-shell states. Second, the kinematical configuration of the loop diagram projects the vertex $V(p_1,-p_1,p_2,-p_2)$  onto the forward limit of the on-shell amplitude $A(s,t)$ evaluated in Euclidean signature,
\eq{
\Delta F &= - \frac{1}{8} V  \int_{p_1} \int_{p_2}  \frac{A( (p_1+p_2)^2,0)}{p^2_1 {p_2}^2} \,  .
}{DeltaF_intermediate}
Since massless force carriers invariably introduce a forward divergence, we avoid this complication by assuming that such interactions are absent.   Furthermore, given a perturbative ultraviolet completion, the tree-level amplitude admits  an analytic low-energy expansion,
\eq{
A(s,0) = \sum_{k=0}^\infty c_{2k} s^{2k} \, ,
}{Apos}
where $c_{2k}$ are Wilson coefficients. 
 Because the amplitude is crossing symmetric, in the $t=0$ forward limit this expression is even in $s$. As is well-known,  $c_{2k} \geq 0$ under the very general assumptions of Lorentz invariance, unitarity, analyticity, and sufficiently convergent ultraviolet behavior, $\lim_{s\rightarrow\infty} A(s,0)/s^{2k} = 0$ \cite{Adams:2006sv}.   The latter is ensured here for $k>0$ due to the Froissart bound \cite{Froissart:1961ux,Martin:1962rt} on unitary theories.  Only free theories can saturate these bounds, so interactions require the strict inequality $c_{2k}>0$.

The integral in \Eq{DeltaF_intermediate} can be rewritten as twice the Boltzmann-weighted phase space integral,
\eq{
\int_p \frac{(\cdots)}{p^2} &=   
 2\int\frac{ d^{D-1} \vec{p}}{(2\pi)^{D-1} 2 |\vec p|}  
\frac{(\cdots)}{e^{|\vec p|/T}-1} =2\int d\Pi (\cdots) \, ,
}{}
yielding \Eq{DeltaP}, which then evaluates to
\eq{
\Delta P &=  \sum_k c_{2k}d_{2k}T^{2D-4+4k} \, .
}{free_energy_long}
Here we have defined the positive constants,
\eq{
&d_{2k} =\tfrac{1}{2}  \int\tfrac{ d^{D-1} \vec{x_1}}{(2\pi)^{D-1} 2 |\vec x_1|} \tfrac{ d^{D-1} \vec{x_2}}{(2\pi)^{D-1} 2 |\vec x_2|} 
\tfrac{ ( 2 |\vec x_1||\vec x_2| - 2\vec x_1\cdot \vec x_2 )^{2k}}{(e^{|\vec x_1|}-1)(e^{|\vec x_2|}-1)}\\
&= 2^{4k-2} \pi^{1/2} 
\tfrac{\Gamma( { 2k+D/2-1})\Gamma(D+2k-2)\zeta(D+2k-2)^2}{(2\pi)^D\Gamma(\frac{D-1}{2})}  >0 \, ,
}{d2k}
where $\vec x_1= \vec p_1/T$ and $\vec x_2= \vec p_2/T$.  Identifying 
\eq{ 
\xi_k = c_{2k} d_{2k} \,, 
}{} 
we then use the fact that
 $c_{2k} ,d_{2k} > 0$ to derive the thermal positivity bound in \Eq{bound}.  In summary, the low-temperature expansion of the pressure, or negative free energy density, exhibits strictly positive coefficients $\xi_k$ in front of each factor of $T^{2D-4+4k}$.

It will be useful to recapitulate the conditions required for the validity of the thermal positivity bound in \Eq{bound}.  First and foremost is the assumption of Lorentz invariance and unitarity, which were crucial for deducing that $c_{2k}>0$ for all $k>0$ in the forward amplitude defined in \Eq{Apos}.  Notably, unitarity does not forbid a dispersive boundary term in the amplitude at $k=0$, which is why this case is unconstrained. In principle these bounds may fail in examples that violate Lorentz invariance, even spontaneously.   A second important assumption is that the two-loop vacuum diagram is dominant, or equivalently that higher-loop and higher-multiplicity contributions are subleading.  As discussed in the appendix, this condition is satisfied in any perturbative ultraviolet completion.  While there has been recent progress on positivity bounds for multiparticle scattering \cite{Arkani-Hamed:2023jwn,Cheung:2025nhw,Elvang:2026pmc,Basile:2026gnd,Cheung:2026lpv}, we will not consider those contributions any further here.

Note that we have not included a mass $m$ for the relativistic bosons of the effective field theory.  Obviously, for $m \gg T$ the thermal fluctuations of these states are exponentially suppressed, while the opposite regime of $ m \ll T $ induces minute corrections. For example, consider an effective field theory of a massless boson whose interactions are all set by an ultraviolet scale $M$.  Then in the presence of a perturbatively small mass $m$, any given low-temperature correction receives a new contribution down by powers of $m^2/T^2$. This effectively bleeds into a more leading term in the temperature expansion with an effective suppression of $m^2/M^2$, which is tiny.

\medskip
\mysec{Examples}Our thermal positivity bounds are instantiated in concrete examples.   These include theories of scalars and vectors with and without flavor.

 \medskip
\noindent {\it Theories of a Single Real Scalar.} The interaction Lagrangian of $\phi^4$ theory is
\be
\Delta L = -\lambda \phi^4 \, .
\ee
The  amplitude gives $c_0= - 24 \lambda$, so the pressure is
\eq{
\Delta P &= -\frac{3\lambda T^{2D-4}}{4\pi^D} \Gamma( D/2-1)^2 \zeta(D-2)^2
\overset{D=4}{\rightarrow} - \frac{\lambda T^4}{48}\, ,
}{}
in agreement with the results of \cite{Kapusta:2006pm,Fernandez-Sarmiento:2025tpy}.  Here positivity would require $\lambda <0$, which strangely implies an unstable potential.  However, one should remember that this contribution corresponds to $k=0$ and our arguments are only robust for $k>0$.  Generalizing to $k=0$ would need the much stronger condition $\lim_{s\rightarrow \infty}A(s,0)\rightarrow0$.  In this case, the unsubtracted dispersion relation for the amplitude has a vanishing boundary term and the quartic coupling is required to be negative.

Next, let us consider a derivatively coupled scalar.  For $(\partial\phi)^4$ theory, the interaction Lagrangian is  
\be 
\Delta L = \lambda(\partial\phi)^4 \, ,
\ee
while the corresponding amplitude,
\be
A(s,t,u) = 2 \lambda (s^2+t^2+u^2) \, ,
\ee
yields $c_2 = 4 \lambda$, so the pressure is
\eq{
\Delta P &=
\frac{16\lambda T^{2D}}{2^D \pi^{D-1/2}}   \frac{\Gamma(D/2+1)\Gamma(D) \zeta(D)^2}{\Gamma(D/2-1/2)}
\overset{D=4}{\rightarrow}\frac{2 \pi^4 \lambda T^8 }{675} \, ,
}{}
which agrees with \cite{Hofmann:2016uzl,Fernandez-Sarmiento:2025tpy}
and is positive.

\medskip
\noindent {\it Theories with Flavor or Spin.} Until now we have considered the simple case of a single real scalar.  Our results generalize to theories with flavor or spin, where the thermal loop diagram depends on the on-shell amplitude {\it summed over all quantum numbers}, so
\eq{
A(s,0)
=
\sum_{i_1,i_2}
\mathcal A_{i_1 i_2 i_2^*i_1^*}(s,0)
=
\sum_{k=0}^\infty c_{2k}s^{2k}\, .
}{}
Here $i_1$ and $i_2$ run over the physical polarizations and flavors of the states in a convention where the external legs are cyclically ordered and all in.    The perturbative correction to the pressure is unchanged from \Eq{free_energy}.

For example, consider the nonlinear sigma model (NLSM).  The corresponding four-point amplitude is
$A_{abcd} (s,t)= R_{abcd} u + R_{acbd}s$, where the indices denote flavor and $R_{abcd}$ is the Riemann curvature tensor in field space.  
The forward limit traces over flavors to produce the vanishing quantity
$\delta_{bc}\delta_{ad} A_{abcd} (s,0) = \delta_{bc}\delta_{ad} (R_{acbd}-R_{abcd})s = 0$.  This is reasonable because the leading NLSM interactions enter at ${\cal O}(p^2)$ and do not contribute to the two-loop free energy. 

In contrast, the ${\cal O}(p^4)$ operators of the NLSM do contribute.  For example, in $D=4$  \cite{Gerber:1988tt,Kapusta:2006pm}, 
the two-loop ${\cal O}(T^8)$ contribution is indeed positive. Meanwhile, for the NLSM in $D=3$ \cite{Hofmann:2009ru,Fernandez-Sarmiento:2025tpy}, 
 the free energy was computed up to ${\cal O}(T^5)$, corresponding to a three-loop  diagram, whereas the ${\cal O}(T^6)$ coefficients would need to be computed to test our bound.

Another interesting example is the Euler-Heisenberg Lagrangian in $D=4$, which at leading order is,
\eq{
&\Delta L ={\lambda_1}(F_{\mu\nu}F^{\mu\nu})^2+{\lambda_2}(F_{\mu\nu}\widetilde F^{\mu\nu})^2 \, ,
}{}
assuming parity conservation.
The helicity amplitudes relevant to the forward trace are \cite{Rebhan:2017zdx}
\eq{
A_{++--}(s,t)&=8(\lambda_1+\lambda_2)s^2\\
A_{+-+-}(s,t)&=8(\lambda_1+\lambda_2)u^2 \, . 
}{}
Summing helicity configurations in the $t=0$ limit yields
\eq{
A(s,0)=32(\lambda_1+\lambda_2)s^2 \, .
}{}
Using $d_2=\pi^4/1350$ in $D=4$, the pressure correction is
\eq{
\Delta P =
{16\pi^4  \over 675} (\lambda_1+\lambda_2)  T^8 \, ,
}{}
in agreement with \cite{Fernandez-Sarmiento:2025tpy}. 
For quantum electrodynamics, we have that $\lambda_1=\alpha^2/90M^4$ and $\lambda_2=7\alpha^2/360M^4$, so
\eq{
\Delta P
=
{22\pi^4\over 30375}\frac{\alpha^2}{M^4}T^8 \, ,
}{}
which matches the two-loop thermal calculation of \cite{Partovi:1993fq,Ravndal:1997wg,Kong:1998ic} and is again positive.

Notably, much work on positivity bounds in scattering has centered on Wilson coefficients in gravitational effective field theories \cite{Camanho:2014apa,Bellazzini:2015cra,Cheung:2016wjt,Tokuda:2020mlf,Caron-Huot:2021rmr,Arkani-Hamed:2021ajd,Caron-Huot:2022ugt,Caron-Huot:2022jli}.  While our thermal positivity bounds can likely be generalized to this context, we leave this possibility for future work.

\medskip
\noindent {\it Theories at Finite Density.}  Many thermodynamic systems of interest exhibit finite density.  In such cases the corresponding chemical potential $\mu$  breaks Lorentz invariance---either explicitly or spontaneously, depending on one's point of view.  This renders the above arguments invalid.  However, for theories that accommodate a smooth $\mu \rightarrow 0$ limit, one expects that the constraint on the Wilson coefficients  $c_{2k} \geq 0$ will still enforce thermal positivity up to perturbatively small $\mu/T$ corrections.

For Lorentz violating theories, crossing symmetry is broken and so the forward amplitude will generically contain nontrivial contributions at odd powers of $s$, defined here to be the squared center-of-mass energy of the non-relativistic system. Interestingly, as argued in \cite{Hui:2023pxc}, for perturbatively small $\mu$ one expects the odd Wilson coefficients $c_{2k-1}$ to share the same sign.  For the superfluid model considered there, these coefficients were all negative but this feature may be model dependent.

\medskip
\mysec{Discussion}In this paper we have derived rigorous bounds on the low-temperature expansion of the pressure, or negative free energy density.  Our results are valid for any Lorentz invariant and unitary effective field theory that is free of long-range forces and whose ultraviolet completion is perturbative.  Under these conditions, we showed that the leading corrections from interactions to the pressure, or negative free energy density, are positive for contributions of the form $T^{2D-4+4k}$ for $k>0$ in $D$-dimensional spacetime.

Our results invite numerous avenues for future inquiry.    For instance, scattering in the absence of Lorentz invariance is not yet fully understood.   Work in the area has focused on theories with a chemical potential, as for example in \cite{Hui:2023pxc,Creminelli:2023kze}.  Anything gleaned from the positivity of Wilson coefficients in a Lorentz-violating effective field theory \cite{Grall:2021xxm,Freytsis:2022aho,Creminelli:2022onn,Creminelli:2023kze,Creminelli:2024lhd} could then be readily imported into the context of thermodynamic observables.  More broadly, a better understanding of positivity for theories in curved spacetime, for example in the background of a black hole or in maximally symmetric spacetimes, would also imply universal signs for thermodynamic quantities.

For Lorentz invariant theories, it is well-known that many positivity properties extend far beyond forward regime \cite{Arkani-Hamed:2020blm,Caron-Huot:2020cmc}.  A natural question then arises: is there a thermodynamic quantity analogous to \Eq{DeltaP} that is expressed in terms of the {\it non-forward} amplitude?  If so, what is its physical interpretation?  We leave this puzzle for future work.

\bigskip
\noindent {\it Acknowledgements}: 
We would like to thank Alex Homrich for early conversations regarding Ref.~\cite{Dashen:1969ep}, and helpful comments from Luca Delacr\'{e}taz, Lucas Fern\'{a}ndez-Sarmiento, Riccardo Penco, and Grant Remmen.  C.C. is supported by the US Department of Energy (Grant No.~DE-SC0011632), the Walter Burke Institute for Theoretical Physics, and the Leinweber Forum for Theoretical Physics.   R.A.R. is supported by US Department of Energy (Grant No.~DE-SC0010118).  R.A.R. would like to thank the Walter Burke Institute for Theoretical Physics and the Leinweber Forum for Theoretical Physics for their hospitality during this work.

\medskip
\mysec{Appendix}Perturbativity of an ultraviolet completion imparts a rigid structure on the higher-dimension operators that appear in the corresponding low-energy theory.  To understand why,  consider an ultraviolet completion with a single heavy mass scale $M$ and a dimensionless weak coupling $g \ll1$.  Since the theory is weakly coupled, perturbation theory is trustworthy and the dynamics are properly captured by tree-level calculations.   

Consider a tree-level $n$-point scattering amplitude in $D$-dimensional spacetime.  By dimensional analysis, we know that $[A_n] =n(1-D/2)+D$.  If this amplitude scales as $k$ overall powers of momentum $p$, then 
\eq{
A_n \sim g^{n-2} p^k M^{n(1-D/2)+D -k} \, .
}{amp_weak_coupling}
Here the factor of $g^{n-2}$ is obvious in the case where the contributing tree diagram is constructed purely from cubic interactions proportional to $g$.  In our conventions, quartic contact interactions will be interpreted as scaling with the same power in the weak coupling constant as a double cubic exchange.

In the low-energy limit the amplitude in \Eq{amp_weak_coupling} induces the effective $n$-point operator,
\eq{
{\cal O }_n  \sim \frac{M^D}{g^2}\left(\frac{\partial}{M}\right)^k \left(\frac{g \phi}{M^{D/2-1}}\right)^n \, .
}{op_weak_coupling}
This scaling accords with usual intuition from weakly coupled ultraviolet completions.  In particular, additional derivatives appear with powers of the inverse gap while additional fields appear with powers of the inverse gap times a weak coupling factor.  For instance, in string theory, each graviton enters with the inverse Planck scale $1/M_{\rm Planck}$  while each derivative enters with inverse string scale $1/M_{\rm string}$.  Similarly in the electroweak model, each longitudinal gauge boson enters with an inverse Higgs vacuum expectation value $1/v$ while each derivative enters with the inverse Higgs mass $1/M_{\rm Higgs}$.

Since $g\ll1$, processes involving more fields rather than fewer fields will be subleading at weak coupling.   Thus, any effect that includes higher-multiplicity tree-level amplitudes will be subleading in $g$. Furthermore, all loop corrections, which necessarily entail additional internal propagating fields, will be subleading in $g$.  In contrast, processes simply involving more derivatives incur no weak coupling penalty.

Now consider a thermal vacuum diagram at some arbitrary loop order.  Our setup only depends on two physical scales, which are the temperature $T$ and the mass $M$.    Hence, up to some overall power of $T$, the diagram only depends on $T/M$ and $g$.  Next, we focus on any internal line in this loop diagram.  By definition, this line terminates at two vertices, each of which pays a single power of $g$, in accordance with \Eq{op_weak_coupling}.  So if we {\it remove} this line to obtain a different diagram with different operator insertions and topology, this new contribution will be less suppressed by $g$ and thus formally more leading in the weak coupling expansion.  We can use this procedure to prune away as many loops as possible from a given diagram.  We then conclude that the lowest possible order in the loop expansion is leading in $g$.  Meanwhile, any arbitrary number of derivative interactions can be added, merely at the cost of factors of $T/M$.  

Thus, in the low-temperature expansion of the pressure correction in \Eq{free_energy}, the coefficient $\xi_k$ corresponding to $T^{2D-4+4k}$  is controlled by the two-loop thermal vacuum diagram, with higher-loop and higher-multiplicity contributions subleading.

To illustrate this power counting in a concrete example, let us consider the NLSM in $D=4$.  The thermal vacuum diagrams induced by various operators are:
\eq{
\begin{tikzpicture}[thick, scale=0.62]

  \tikzset{
    vertex/.style={
      draw=black,
      fill=gray!20,
      circle,
      minimum size=8pt,
      inner sep=0pt
    }
  }

  \def\xlabel{-2.35}   
  \def\xeqone{2.8}
  \def\xA{4.4}

  \node[anchor=east] at (\xlabel,0)
    {$\dfrac{g^2}{M^4}\,\partial^4 \phi^4 \quad \rightarrow$};

  \begin{scope}[shift={(0,0)}]
    \draw (0,0.95) circle (0.95);
    \draw (0,-0.95) circle (0.95);
    \node[vertex] at (0,0) {};
  \end{scope}

  \node at (\xeqone,0) {$\simeq$};
  \node at (\xA,0) {$\dfrac{g^2 T^8}{M^4}$};

  \node[anchor=east] at (\xlabel,-4.35)
    {$\dfrac{g^2}{M^2}\,\partial^2 \phi^4 \quad \rightarrow$};

  \begin{scope}[shift={(0,-4.35)}]
    \coordinate (L) at (-1.35,0);
    \coordinate (R) at (1.35,0);

    \draw (L) .. controls (-1.00,1.85) and (1.00,1.85) .. (R);
    \draw (L) .. controls (-0.35,0.70) and (0.35,0.70) .. (R);
    \draw (L) .. controls (-0.35,-0.70) and (0.35,-0.70) .. (R);
    \draw (L) .. controls (-1.00,-1.85) and (1.00,-1.85) .. (R);

    \node[vertex] at (L) {};
    \node[vertex] at (R) {};
  \end{scope}

  \node at (\xeqone,-4.35) {$\simeq$};
  \node at (\xA,-4.35) {$\dfrac{g^4 T^8}{M^4}$};

  \node[anchor=east] at (\xlabel,-8.7)
    {$\dfrac{g^4}{M^4}\,\partial^2 \phi^6 \quad \rightarrow$};

  \begin{scope}[shift={(0,-8.7)}]
    \foreach \ang in {90,210,330} {
      \begin{scope}[rotate=\ang]
        \draw (0,0)
          .. controls (0.38,0.22) and (0.85,0.72) .. (0.62,1.18)
          .. controls (0.30,1.55) and (-0.30,1.55) .. (-0.62,1.18)
          .. controls (-0.85,0.72) and (-0.38,0.22) .. (0,0);
      \end{scope}
    }
    \node[vertex] at (0,0) {};
  \end{scope}

  \node at (\xeqone,-8.7) {$\simeq$};
  \node at (\xA,-8.7) {$\dfrac{g^4 T^8}{M^4}$};

\end{tikzpicture}
}{}
Here we have written each operator together with its suppression in the perturbative coupling mandated by \Eq{op_weak_coupling}.  The contributions from higher loops or higher multiplicity are subdominant in the weak coupling expansion.  In particular, the two-loop contribution from a single insertion of $\partial^4 \phi^4$ is leading.  Note that the analogous two-loop contribution from a single insertion of $\partial^2 \phi^4$ is zero.  Last but not least, any operators of similar form with additional powers of derivatives will simply enter with $T/M$ suppression.

\bibliographystyle{apsrev4-1}
\bibliography{Thermal_Positivity}

\end{document}